\documentclass[aps,prc,superscriptaddress,twoside,twocolumn,showpacs]{revtex4}
\usepackage{amsmath,amssymb}
\usepackage{graphicx}
\usepackage{graphics}
\usepackage{epsfig}
\usepackage{subfigure}

\newcommand{\pol}[1]{\mathaccent"017E{#1}}
\def\fmn#1#2{\mbox{${\textstyle \frac{#1}{#2}}$}}
\newcommand{\Szero}{\mbox{$^{1\!}S_0$}}
\def\fmn#1#2{\mbox{${\textstyle \frac{#1}{#2}}$}}
\newcommand{\pppi}{\mbox{$pp\to \{pp\}_{\!s\,}\pi^0$}}
\newcommand{\vpppi}{\mbox{$\pol{p}p\to \{pp\}_{\!s\,}\pi^0$}}

\newcommand{\npdpi}{\mbox{$np\to d\pi^0$}}
\newcommand{\vnpdpi}{\mbox{$\pol{n}p\to d\pi^0$}}
\newcommand{\nppppi}{\mbox{$np\to \{pp\}_{\!s}\pi^-$}}
\newcommand{\pnpppi}{\mbox{$pn\to \{pp\}_{\!s}\pi^-$}}
\newcommand{\vpnpppi}{\mbox{$\pol{p}n\to \{pp\}_{\!s}\pi^-$}}
\newcommand{\vnvpppppi}{\mbox{$\pol{n}\,\pol{p}\to \{pp\}_{\!s}\pi^-$}}

\newcommand{\half}{\mbox{${\textstyle \frac{1}{2}}$}}           

\begin{document}
\title{Measurement of spin observables in the quasi-free
$\boldsymbol{np\to \{pp\}_{\!s}\pi^-}$ reaction at 353~MeV}
\author{S.~Dymov}\email{s.dymov@fz-juelich.de}
\affiliation{Physikalisches Institut II, Universit{\"a}t Erlangen-N{\"u}rnberg, D-91058 Erlangen, Germany}
\affiliation{Laboratory of Nuclear Problems, Joint Institute for Nuclear Research, RU-141980 Dubna, Russia}
\author{V.~Shmakova}
\affiliation{Laboratory of Nuclear Problems, Joint Institute for Nuclear
  Research, RU-141980 Dubna, Russia}
\affiliation{Institut f\"ur Kernphysik and J\"ulich Centre for Hadron
Physics, Forschungszentrum J\"ulich, D-52425 J\"ulich, Germany}

\author{T.~Azaryan}
\affiliation{Laboratory of Nuclear Problems, Joint Institute for Nuclear
  Research, RU-141980 Dubna, Russia}

\author{S.~Barsov}
\affiliation{St. Petersburg Nuclear Physics Institute, RU-188350 Gatchina,
  Russia}

\author{V.~Baru}
\affiliation{Institute of Theoretical Physics, Ruhr-Universit\"{a}t, D-44780
Bochum, Germany}
\affiliation{Institute for Theoretical and Experimental Physics,  RU-117218 Moscow,
Russia}

\author{P.~Benati}
\affiliation{Universit\`a di Ferrara and INFN, IT-44100 Ferrara,
Italy}


\author{D.~Chiladze}
\affiliation{Institut f\"ur Kernphysik and J\"ulich Centre for Hadron
Physics, Forschungszentrum J\"ulich, D-52425 J\"ulich, Germany}
\affiliation{High Energy Physics Institute, Tbilisi State University, GE-0186
Tbilisi, Georgia}

\author{A.~Dzyuba}
\affiliation{St. Petersburg Nuclear Physics Institute, RU-188350 Gatchina,
  Russia}

\author{R.~Engels}
\affiliation{Institut f\"ur Kernphysik and J\"ulich Centre for Hadron
Physics, Forschungszentrum J\"ulich, D-52425 J\"ulich, Germany}

\author{M.~Gaisser}
\affiliation{Institut f\"ur Kernphysik and J\"ulich Centre for Hadron
Physics, Forschungszentrum J\"ulich, D-52425 J\"ulich, Germany}

\author{R.~Gebel}
\affiliation{Institut f\"ur Kernphysik and J\"ulich Centre for Hadron
Physics, Forschungszentrum J\"ulich, D-52425 J\"ulich, Germany}

\author{K.~Grigoryev}
\affiliation{Institut f\"ur Kernphysik and J\"ulich Centre for Hadron
Physics, Forschungszentrum J\"ulich, D-52425 J\"ulich, Germany}
\affiliation{St. Petersburg Nuclear Physics Institute, RU-188350 Gatchina,
  Russia}

\author{P.~Goslawski}
\affiliation{Institut f\"ur Kernphysik, Universit\"at M\"unster,
D-48149 M\"unster, Germany}

\author{G.~Guidoboni}
\affiliation{Universit\`a di Ferrara and INFN, IT-44100 Ferrara,
Italy}


\author{M.~Hartmann}
\affiliation{Institut f\"ur Kernphysik and J\"ulich Centre for Hadron
Physics, Forschungszentrum J\"ulich, D-52425 J\"ulich, Germany}

\author{A.~Kacharava}\email{a.kacharava@fz-juelich.de}
\affiliation{Institut f\"ur Kernphysik and J\"ulich Centre for Hadron
Physics, Forschungszentrum J\"ulich, D-52425 J\"ulich, Germany}

\author{V.~Kamerdzhiev}
\affiliation{Institut f\"ur Kernphysik and J\"ulich Centre for Hadron
Physics, Forschungszentrum J\"ulich, D-52425 J\"ulich, Germany}

\author{A.~Khoukaz}
\affiliation{Institut f\"ur Kernphysik, Universit\"at M\"unster, D-48149 M\"unster, Germany}

\author{V.~Komarov}
\affiliation{Laboratory of Nuclear Problems, Joint Institute for Nuclear
  Research, RU-141980 Dubna, Russia}

\author{P.~Kulessa}
\affiliation{Institute of Nuclear Physics, PL-31342 Cracow,
Poland}

\author{A.~Kulikov}
\affiliation{Laboratory of Nuclear Problems, Joint Institute for Nuclear
  Research, RU-141980 Dubna, Russia}

\author{V.~Kurbatov}
\affiliation{Laboratory of Nuclear Problems, Joint Institute for Nuclear
  Research, RU-141980 Dubna, Russia}

\author{A.~Lehrach}
\affiliation{Institut f\"ur Kernphysik and J\"ulich Centre for Hadron
Physics, Forschungszentrum J\"ulich, D-52425 J\"ulich, Germany}

\author{P.~Lenisa}
\affiliation{Universit\`a di Ferrara and INFN, IT-44100 Ferrara,
Italy}

\author{V.~Lensky}
\affiliation{School of Physics and Astronomy, University of Manchester, Manchester M13 9PL,
UK}
\affiliation{Institute for Theoretical and Experimental Physics,  RU-117218 Moscow,
Russia}

\author{N.~Lomidze}
\affiliation{High Energy Physics Institute, Tbilisi State University, GE-0186
Tbilisi, Georgia}

\author{B.~Lorentz}
\affiliation{Institut f\"ur Kernphysik and J\"ulich Centre for Hadron
Physics, Forschungszentrum J\"ulich, D-52425 J\"ulich, Germany}

\author{G.~Macharashvili}
\affiliation{Laboratory of Nuclear Problems, Joint Institute for Nuclear
  Research, RU-141980 Dubna, Russia}
\affiliation{High Energy Physics Institute, Tbilisi State University, GE-0186
Tbilisi, Georgia}

\author{R.~Maier}
\affiliation{Institut f\"ur Kernphysik and J\"ulich Centre for Hadron
Physics, Forschungszentrum J\"ulich, D-52425 J\"ulich, Germany}

\author{D.~Mchedlishvili}
\affiliation{Institut f\"ur Kernphysik and J\"ulich Centre for Hadron
Physics, Forschungszentrum J\"ulich, D-52425 J\"ulich, Germany}
\affiliation{High Energy Physics Institute, Tbilisi State University, GE-0186
Tbilisi, Georgia}

\author{S.~Merzliakov}
\affiliation{Laboratory of Nuclear Problems, Joint Institute for Nuclear
  Research, RU-141980 Dubna, Russia}
\affiliation{Institut f\"ur Kernphysik and J\"ulich Centre for Hadron
Physics, Forschungszentrum J\"ulich, D-52425 J\"ulich, Germany}

\author{M.~Mielke}
\affiliation{Institut f\"ur Kernphysik, Universit\"at M\"unster,
  D-48149 M\"unster, Germany}

\author{M.~Mikirtychyants}
\affiliation{Institut f\"ur Kernphysik and J\"ulich Centre for Hadron
Physics, Forschungszentrum J\"ulich, D-52425 J\"ulich, Germany}
\affiliation{St. Petersburg Nuclear Physics Institute, RU-188350 Gatchina,
  Russia}

\author{S.~Mikirtytchiants}
\affiliation{Institut f\"ur Kernphysik and J\"ulich Centre for Hadron
Physics, Forschungszentrum J\"ulich, D-52425 J\"ulich, Germany}
\affiliation{St. Petersburg Nuclear Physics Institute, RU-188350 Gatchina,
  Russia}

\author{M.~Nioradze}
\affiliation{High Energy Physics Institute, Tbilisi State University, GE-0186
Tbilisi, Georgia}

\author{D.~Oellers}
\affiliation{Institut f\"ur Kernphysik and J\"ulich Centre for Hadron
Physics, Forschungszentrum J\"ulich, D-52425 J\"ulich, Germany}

\author{H.~Ohm}
\affiliation{Institut f\"ur Kernphysik and J\"ulich Centre for Hadron
Physics, Forschungszentrum J\"ulich, D-52425 J\"ulich, Germany}

\author{A.~Polyanskiy}
\affiliation{Institute for Theoretical and Experimental Physics,  RU-117218 Moscow,
Russia}

\author{M.~Papenbrock}
\affiliation{Institut f\"ur Kernphysik, Universit\"at M\"unster,
D-48149 M\"unster, Germany}

\author{D.~Prasuhn}
\affiliation{Institut f\"ur Kernphysik and J\"ulich Centre for Hadron
Physics, Forschungszentrum J\"ulich, D-52425 J\"ulich, Germany}

\author{F.~Rathmann}
\affiliation{Institut f\"ur Kernphysik and J\"ulich Centre for Hadron
Physics, Forschungszentrum J\"ulich, D-52425 J\"ulich, Germany}

\author{V.~Serdyuk}
\affiliation{Laboratory of Nuclear Problems, Joint Institute for Nuclear
  Research, RU-141980 Dubna, Russia}
\affiliation{Institut f\"ur Kernphysik and J\"ulich Centre for Hadron
Physics, Forschungszentrum J\"ulich, D-52425 J\"ulich, Germany}

\author{H.~Seyfarth}
\affiliation{Institut f\"ur Kernphysik and J\"ulich Centre for Hadron
Physics, Forschungszentrum J\"ulich, D-52425 J\"ulich, Germany}

\author{E.~Steffens}
\affiliation{Physikalisches Institut II, Universit{\"a}t Erlangen-N{\"u}rnberg, D-91058 Erlangen, Germany}

\author{H.J.~Stein}
\affiliation{Institut f\"ur Kernphysik and J\"ulich Centre for Hadron
Physics, Forschungszentrum J\"ulich, D-52425 J\"ulich, Germany}

\author{H.~Stockhorst}
\affiliation{Institut f\"ur Kernphysik and J\"ulich Centre for Hadron
Physics, Forschungszentrum J\"ulich, D-52425 J\"ulich, Germany}

\author{H.~Str\"oher}
\affiliation{Institut f\"ur Kernphysik and J\"ulich Centre for Hadron
Physics, Forschungszentrum J\"ulich, D-52425 J\"ulich, Germany}

\author{M.~Tabidze}
\affiliation{High Energy Physics Institute, Tbilisi State University, GE-0186
Tbilisi, Georgia}

\author{S.~Trusov}
\affiliation{Institut f\"ur Kern- und Hadronenphysik,
Forschungszentrum Rossendorf, D-01314 Dresden, Germany}

\author{D.~Tsirkov}
\affiliation{Laboratory of Nuclear Problems, Joint Institute for Nuclear
  Research, RU-141980 Dubna, Russia}

\author{Yu.~Uzikov}
\affiliation{Laboratory of Nuclear Problems, Joint Institute for Nuclear
  Research, RU-141980 Dubna, Russia}

\author{Yu.~Valdau}
\affiliation{Institut f\"ur Kernphysik and J\"ulich Centre for Hadron
Physics, Forschungszentrum J\"ulich, D-52425 J\"ulich, Germany}
\affiliation{St. Petersburg Nuclear Physics Institute, RU-188350 Gatchina,
  Russia}

\author{Ch.~Weidemann}
\affiliation{Institut f\"ur Kernphysik and J\"ulich Centre for Hadron
Physics, Forschungszentrum J\"ulich, D-52425 J\"ulich, Germany}

\author{C.~Wilkin}
\affiliation{Physics and Astronomy Department, UCL, Gower Street, London WC1E
6BT, United Kingdom}

\author{P.~ W\"ustner}
\affiliation{Zentralinstitut f\"ur Elektronik, Forschungszentrum J\"ulich, D-52425 J\"ulich, Germany}

\author{Q.~J.~Ye}
\affiliation{Institut f\"ur Kernphysik and J\"ulich Centre for Hadron
Physics, Forschungszentrum J\"ulich, D-52425 J\"ulich, Germany}
\affiliation{Department of Physics and Triangle Universities Nuclear Laboratory,
 Duke University, Durham, NC 27708, USA}

\author{M.~Zhabitsky}
\affiliation{Laboratory of Nuclear Problems, Joint Institute for Nuclear Research, RU-141980 Dubna, Russia}

\date{\today}
\begin{abstract} The transverse spin correlations $A_{x,x}$ and
$A_{y,y}$ have been measured in the $\pol{d}\pol{p}\to
p_{\text{spec}}\{pp\}_{\!s}\pi^-$ reaction at COSY-ANKE at 353~MeV per nucleon. Here
$\{pp\}_{\!s}$ denotes a proton-proton pair with low excitation energy, which
is dominantly in the \Szero\ state. By measuring three protons in the final
state it was possible to extract events where there was a spectator proton
$p_{\text{spec}}$ so that the reaction could be interpreted in terms of
quasi-free \vnvpppppi. The proton analyzing power in this reaction was also
deduced from this data set by averaging over the polarization of the deuteron
beam. The values of $A_y^p$ were shown to be consistent with a refined
analysis of our earlier results obtained with a polarized proton incident on
a deuterium target. Taking these data in combination with our previous
measurements of the differential cross sections and analyzing powers in the
\vpppi\ reaction, a more robust partial wave decomposition was achieved.
Three different acceptable solutions were found and the only way of resolving
this ambiguity without further theoretical input would be through a
measurement of the mixed spin-correlation parameter $A_{x,z}$.
\end{abstract}
\maketitle

%
%

\section{Introduction}
\label{sec1}

One of the major challenges in today's physics is to relate the properties of
few– nucleon systems and nuclei to the theory of strong interactions, QCD. In
this respect there has been significant theoretical progress in establishing
an effective field theory that, while having a clear cut connection to QCD,
allows one to study processes involving strongly interacting particles within
a well defined perturbative scheme. It is chiral symmetry that provides the
preconditions for the construction of an effective field theory, called
chiral perturbation theory or simply $\chi$PT~\cite{WEI1979,GAS1984}.

A modification to the standard $\chi$PT approach is necessary when it is
applied to pion production in nucleon-nucleon collisions. The large scale,
introduced by the initial momentum, has to be considered
explicitly~\cite{HAN2000,HAN2002}. Thus, a proper expansion scheme for pion
production is now established and a high-precision calculation for the
reactions $NN \to NN\pi$ is currently under
way~\cite{LEN2006,FIL2012,BAR2009,FIL2009}.

One important step forward in our understanding of pion reactions at low
energies~\cite{HAN2004} will be to establish that the same short-range $NN
\to NN\pi$ vertex contributes to both $p$-wave pion production and to low
energy three-nucleon scattering, where a crucial role is played by the
identical production operator~\cite{EPE2005,EPE2002}. Apart from pion
production and the three-nucleon force, this short-range operator contributes
to electroweak processes, such as $pp\to de^+\nu_e$, triton $\beta$
decay~\cite{PAR2003,GAR2006a,GAZ2009}, and muon absorption on the deuteron
$\mu^- d \to nn \nu_{\mu}$ ~\cite{AND2010,AND2000,MAR2012,ADA2012,MAR2003}, as well
as to reactions involving photons, \emph{e.g.}, $\pi d \to \gamma
NN$~\cite{GAR2006a,GAR2006} and $\gamma d \to
nn\pi^+$~\cite{LEN2005,LEN2007}. The strength of this production operator
cannot be fixed from processes in the one-nucleon sector, such as in
pion-nucleon scattering~\cite{ORD1992}. The missing term corresponds to an
effective $NN \to NN\pi$ vertex, where the pion is in a $p$-wave and both
initial and final $NN$ pairs are in relative $S$ waves. It is our aim to
extract the relevant partial wave amplitude in pion production from
experiment. This is a precondition for a reliable determination of this
contact term.

The COSY-ANKE collaboration has embarked on an ambitious program of
performing a complete set of measurements of the $NN\to\{pp\}_{\!s\,}\pi$
reactions at low energy so that a full amplitude analysis can be carried
out~\cite{SPIN}. By selecting events with excitation energy in the
proton-proton system $E_{pp} < 3$~MeV, the resulting diproton $\{pp\}_{\!s}$
is overwhelmingly in the \Szero\ state. In this case all the possible
information on the production amplitudes can be obtained by using polarized
beams and targets; no measurements of the polarizations of the final protons
are required.

As parts of this program, we have already reported on measurements at
$T_p=353$~MeV of the cross sections $d\sigma/d\Omega$ and proton analyzing
powers $A_y^p$ in the \vpppi\ reaction~\cite{TSI2012} and the quasi-free
\vpnpppi\ reaction, where a polarized proton beam was incident on a deuterium
target~\cite{DYM2012}. We here complement this information through
measurements in inverse kinematics, with a polarized deuteron beam colliding
with a polarized hydrogen target. This leads to a determination of the
proton-neutron transverse spin correlations $A_{x,x}$ and $A_{y,y}$ in the
quasi-free \vnvpppppi\ reaction at the same energy and also an independent
measurement of the proton analyzing power for $n\pol{p}\to\{pp\}_{\!s}\pi^-$.
When this information is used in conjunction with a refined analysis of
the earlier data, a more robust amplitude decomposition can be developed
for both the $I=1$ and $I=0$ channels. The residual ambiguities in this
procedure can also be clearly displayed.

The approach to pion production described here was first initiated at TRIUMF
through pioneering measurements of the \pnpppi\ differential cross
section~\cite{HAH1999} and the proton analyzing power in the \vpnpppi\
reaction~\cite{DUN1998} at 353~MeV and it was the existence of these data
that influenced our choice of beam energy. These results were complemented by
TRIUMF data on the quasi-free absorption of a $\pi^-$ on the diproton pair in
the $^3$He nucleus~\cite{HAH1996}. This provided the shape of the
$\pi^-\{pp\}_{\!s}\to pn$ cross section but not the normalization.

The group made a partial wave analysis of their results~\cite{HAH1999} using
a methodology developed earlier~\cite{PIA1986} but their data did not extend
over the whole angular region. A more serious drawback was the lack of
comparable data on the \pppi\ reaction that could be used to constrain the
isospin $I=1$ amplitudes. They therefore ruled out a solution with a large
pion $d$-wave and it was only shortly afterwards that the CELSIUS \pppi\
differential cross section measurements were published that showed that there
were indeed large pion $d$-wave contributions, even at relatively low
energy~\cite{BIL2001}.

The general amplitude structure for the $NN\to\{pp\}_{\!s}\pi$ reaction is
discussed in Sec.~\ref{sec2}, where relations between the amplitudes and the
different possible observables are described. Of especial importance are the
symmetry relations that link the observables. These relations are, of course,
respected by the partial wave development that is also presented here. Since
the data reported in this paper were taken quite near threshold, we only keep
terms up to and including pion $d$-waves. The experimental apparatus and
procedure is the subject of Sec.~\ref{sec3}. A particular concern here
compared to our previous work at ANKE~\cite{TSI2012,DYM2012} is the use of
the polarized gas target cell that was required in order to achieve a viable
luminosity. Naturally, high and well determined beam and target polarizations
are critical in any measurement of a spin correlation but, because the
differential cross section had already been measured~\cite{DYM2012}, an
absolute normalisation was of much lesser importance.

Section~\ref{sec4} is devoted to the treatment of the data taken in
deuteron-proton collisions with both beam and target being polarized. Unlike
the spectator proton in the deuterium target work, that in the
$dp\to\{pp\}_{\!s}\pi^-p_{\text{spec}}$ case is fast and is registered in the
forward detector system of the ANKE magnetic spectrometer. Although the
principles of the analysis of the two experiments are similar, they differ
significantly in their details, especially with respect to the kinematics
reconstruction and the handling of the background. The different approaches
to the polarimetry are also discussed here. Before the results of the current
experiment are presented in Sec.~\ref{sec5}, we first explain how a
reanalysis of the deuterium target data of Ref.~\cite{DYM2012} was
achieved by using fully the timing information provided by the apparatus.
This effectively doubled the statistics in the $A_y^p$ measurement. The
values achieved here are consistent with the published results and also with
those derived from the new polarized hydrogen-target experiment. The latter
also led to a consistent shape for the differential cross section. Finally in
this section are presented the results on the spin-correlation coefficients.
The fact that, on symmetry grounds, $A_{y,y}=1$ provided an extra check on
the product of the beam and target polarizations and led to a more stable
evaluation of $A_{x,x}$.

Even after making phase assumptions on the isospin-1 production amplitudes,
the partial wave analysis of Sec.~\ref{sec6} results in three distinct
solutions that have very similar statistical significance. They all reproduce
the measured values of the differential cross section, the proton analyzing
power, and transverse spin correlation for both $NN\to \{pp\}_{\!s}\pi$
reactions. Though one of the solutions might be preferred on theoretical
grounds, as stressed in our conclusions of Sec.~\ref{sec7}, the ambiguities
could only be resolved experimentally through the difficult measurement of
the mixed spin-correlation parameter $A_{x,z}$.

%
%
\section{Amplitudes, Observables, and Partial Waves}
\label{sec2}

\subsection{Polarization observables}

In the frame where the $z$-direction is along the beam and the $y$-direction
is perpendicular to the reaction plane, the differential cross section for
the \pppi\ or \nppppi\ reaction takes the form
\begin{eqnarray}
\nonumber
\frac{d\sigma}{d\Omega}&=&\left(\frac{d\sigma}{d\Omega}\right)_{\!0}
\big[1 + P_{y}A_{y}^P + Q_{y}A_{y}^Q
+P_{y}Q_yA_{y,y}\\
&&\hspace{-1.3cm} +P_{x}Q_xA_{x,x} +P_{z}Q_zA_{z,z}
+P_{x}Q_zA_{x,z}+P_{z}Q_xA_{z,x}\big],
\end{eqnarray}
where $\boldsymbol{P}$ and $\boldsymbol{Q}$ are the beam and target
polarizations, $(d\sigma/d\Omega)_0$ is the unpolarized cross section, and
parity conservation is assumed. The beam $A_y^P$ and target $A_y^Q$ analyzing
powers, as well as the spin-correlation parameters $A_{ij}$, are all
functions of the pion polar angle $\theta_{\pi}$.

When, as in this experiment, the beam and target are both polarized
perpendicular to the plane of the COSY ring with polarizations $P$ and $Q$,
respectively, it is more convenient to rewrite the expression in the fixed
COSY frame as
\begin{eqnarray} \nonumber
\hspace{-1cm}\frac{d\sigma}{d\Omega}=\left(\frac{d\sigma}{d\Omega}\right)_{\!0}
\big[1 \hspace{-4mm}&&+ (PA_y^P + QA_y^Q)\cos\varphi_{\pi}\\
&&+PQ\left(A_{y,y}\cos^2\varphi_{\pi} +
A_{x,x}\sin^2\varphi_{\pi}\right)\big],
\label{asymmetry}
\end{eqnarray}
where $\varphi_{\pi}$ is the azimuthal angle of the pion in
the laboratory reference frame (Fig. \ref{fig:ANKE}).
 We neglect here the small $P_z$ and $Q_z$ components appearing in a quasi-free
measurement due to the Fermi motion in the deuteron. Their effect is included
in the systematic error in the analysis.

\subsection{Production amplitudes and observables}

The spin structure of the \pppi\ or the \nppppi\ reaction is that of
$\half^+\half^+\to 0^+0^-$. Parity and angular momentum conservation then
require that the initial nucleon-nucleon pair has spin $S=1$. There are only
two independent scalar amplitudes, $A$ and $B$, which we define in terms of
the full amplitude $\mathcal{M}$ through
\begin{equation}
\mathcal{M}=\boldsymbol{S}\cdot
\left(A\,\hat{\boldsymbol{p}} + B\,\hat{\boldsymbol{k}}\right),
\label{amps2}\end{equation}%
where $\boldsymbol{S}$ is the polarization vector of the initial spin-triplet
$NN$ state. $\hat{\boldsymbol{p}}$ and $\hat{\boldsymbol{k}}$ are unit
vectors in the center-of-mass (c.m.) frame along the directions of the
incident proton and final pion momenta, respectively.

The possible observables are expressed in terms of the
amplitudes $A$ and $B$ as
\begin{eqnarray}
\nonumber\left(\frac{d\sigma}{d\Omega}\right)_{\!0}&=&\frac{k}{4p}\left(|A|^2+|B|^2
+2\,\textrm{Re}[AB^*]\cos\theta_{\pi}\right)\!,\\
\nonumber
A_y^P\left(\frac{d\sigma}{d\Omega}\right)_{\!0}&=&
\frac{k}{4p}\left(
2\,\textrm{Im}[AB^*]\sin\theta_{\pi}\right)\!,\\
\nonumber
A_{x,x}\left(\frac{d\sigma}{d\Omega}\right)_{\!0}&=&
\frac{k}{4p}\left(|A|^2+|B|^2\cos 2\theta_{\pi}
+2\,\textrm{Re}[AB^*]\cos\theta_{\pi}\right)\!,\\
\nonumber
A_{x,z}\left(\frac{d\sigma}{d\Omega}\right)_{\!0}&=&
-\frac{k}{4p}\left(2|B|^2\sin\theta_{\pi}\cos\theta_{\pi}
+2\,\textrm{Re}[AB^*]\sin\theta_{\pi}\right)\!,\\
\label{observables} A_y^Q=A_y^P,&&\hspace{-3mm} A_{y,y}=1,\ A_{z,z}=-A_{x,x},\
A_{z,x}=A_{x,z}.
\end{eqnarray}
The reaction is treated as a quasi-two-body one and, in the evaluation of the
phase space factor $k/p$, the small range of excitation energies in the
diproton is neglected.

The observables are not all independent and it is straightforward to show
that, for any pion production angle,
\begin{equation}
\big(A_y\big)^2 + \big(A_{x,x}\big)^2 + \big(A_{x,z}\big)^2 =1.
\label{quadratic}
\end{equation}
This means that, if two of the quantities are well measured, there remains
only a sign ambiguity in the determination of the third from this quadratic
relation.

In addition to the spin dependence of the reaction, there are also two
isospin amplitudes $\mathcal{M}^{I=1}$ and $\mathcal{M}^{I=0}$ and, in terms
of these, $\mathcal{M}(\pppi) =\mathcal{M}^{I=1}$ and
$\mathcal{M}(\nppppi)=(\mathcal{M}^{I=1} +\mathcal{M}^{I=0})/\sqrt{2}$. Since
the initial nucleons are in a spin-triplet state, the Pauli principle
requires that $\mathcal{M}^{I=1}$ is antisymmetric under the reflection
$\boldsymbol{p}\to \boldsymbol{-p}$ whereas $\mathcal{M}^{I=0}$ is symmetric.
Due to the presence of the proton momentum factor in Eq.~(\ref{amps2}), this
constraint translates into the
requirements that%
\begin{eqnarray}
\nonumber
A^{I=1}(\cos\theta_{\pi}) &=& \phantom{-}A^{I=1}(-\cos\theta_{\pi}),\\
\nonumber
B^{I=0}(\cos\theta_{\pi}) &=& \phantom{-}B^{I=0}(-\cos\theta_{\pi}),\\
\nonumber
B^{I=1}(\cos\theta_{\pi}) &=& -B^{I=1}(-\cos\theta_{\pi}),\\
\label{symmetries}
A^{I=0}(\cos\theta_{\pi}) &=& -A^{I=0}(-\cos\theta_{\pi}).
\end{eqnarray}

As a consequence of Eq.~(\ref{symmetries}), both $B^{I=1}$ and $A^{I=0}$
vanish at $\theta_{\pi}=90^{\circ}$ and this leads to the important relation
at this particular angle:
\begin{equation}
\big(1+A_{x,x}\big)\frac{d\sigma}{d\Omega}(\nppppi)
= \frac{d\sigma}{d\Omega}(\pppi).
\label{Axx_rel}
\end{equation}
Independent of any assumptions made in the subsequent data analysis, the
value of the spin-correlation $A_{x,x}$ in the \nppppi\ reaction at
$90^{\circ}$ is fixed completely by the unpolarized \pppi\ and \nppppi\
differential cross sections. However, the quasi-free nature of the $\pi^-$
production experiment, as well as the mass differences among both the pions
and nucleons, means that there is uncertainty in the relative normalizations
of the two unpolarized measurements and so the direct study of $A_{x,x}$
presented in the current work is definitely preferable.

A second useful result for the \pppi\ reaction that follows from the symmetry
relations of Eq.~(\ref{symmetries}) is that $A_{x,x}=1$ at $90^{\circ}$.

A complete set of measurements of the observables in
Eq.~(\ref{observables}) would fix the magnitudes of the
amplitudes $A$ and $B$ and their relative phase but the overall
phase, which is a function of $\theta_{\pi}$ is clearly
undetermined. Since this phase function can be different for
$\pi^0$ and $\pi^-$ production, extra assumptions are required
to avoid the consequent ambiguities. These are expressed most
clearly in terms of the partial wave amplitudes, to which we
now turn.

\subsection{Partial wave decomposition}
\label{sec:pwa}
Our earlier experiments~\cite{TSI2012,DYM2012}, and the one
reported here, were carried out in the vicinity of 353~MeV and,
at such a low beam energy, one may expect that very few pion
partial waves will contribute. Keeping terms up to pion $d$
waves, there are three possible transitions from the $I=1$
initial state, viz.\ $^{3\!}P_0\to\, ^{1\!}S_0s$,
$^{3\!}P_2\to\, ^{1\!}S_0d$, and $^{3\!}F_2\to\, ^{1\!}S_0d$
and we denote the corresponding amplitudes by $M_s^P$, $M_d^P$,
and $M_d^F$, respectively. For the $I=0$ state, there are the
two $p$-wave transitions, $^{3\!}S_1\to\, ^{1\!}S_0p$ and
$^{3\!}D_1\to\, ^{1\!}S_0p$, whose amplitudes we call $M_p^S$
and $M_p^D$, respectively.

The scalar amplitudes can be decomposed in terms of these
partial waves as
\begin{eqnarray}
\nonumber
A^{I=1}&=&M_s^P-\fmn{1}{3}M_d^P+M_d^F\left(\cos^2\theta_{\pi}-\fmn{1}{5}\right),\\
\nonumber
B^{I=1}&=&\left(M_d^P-\fmn{2}{5}M_d^F\right)\cos\theta_{\pi},\\
\nonumber
A^{I=0}&=&M_p^D\cos\theta_{\pi},\\
\label{pwamps}
B^{I=0}&=&M_p^S-\fmn{1}{3}M_p^D\,,
\end{eqnarray}
which, of course, respect the symmetries shown in
Eq.~(\ref{symmetries}).

The partial-wave amplitudes, which depend purely on energy,
 have a threshold behaviour like $k^{\ell}$, where
$\ell$ is the pion angular momentum. However, this power counting is slightly
deceptive because it is well known that the $s$-wave amplitude in pion
production is suppressed. As a consequence, it is reasonable for $\pi^0$
production to introduce all the terms of Eq.~(\ref{pwamps}) into the
expressions for the observables in Eq.~(\ref{observables}) since any
arising from $s$-$g$ interference, which have the same threshold dependence
as $d$-$d$ terms, are expected to be very small. This procedure preserves
exactly the quadratic relation of Eq.~(\ref{quadratic}). However, our neglect
of $p$-$f$ interference for $\pi^-$ production may be less justified.

We have already argued~\cite{TSI2012,DYM2012} that,
 even with a measurement of the complete set of
observables in Eq.~(\ref{observables}), there are overall phase uncertainties
that are functions of $\theta_{\pi}$. These are then reflected in ambiguities
in the partial wave amplitudes which can only be resolved by making further
assumptions. For uncoupled partial waves, provided the inelasticity is very
small, the Watson theorem fixes the phase induced by the initial state
interaction to that of the elastic nucleon-nucleon scattering~\cite{WAT1952}.
These conditions apply for the $^{3\!}P_0$ partial wave and so we take
$M_s^P=|M_s^P|e^{i\delta_{^{3\!}P_0}}$, with
$\delta_{^{3\!}P_0}=-14.8^{\circ}$~\cite{ARN2007}. The phase associated with
the $^{1\!}S_0$ final $pp$ state is not included because it is common to all
partial waves and does not influence the observables.

For coupled channels, such as $^{3\!}P_2-\,^{3\!}F_2$, the conditions of the
Watson theorem do not strictly apply. However, phase shift analysis of $pp$
data at 353~MeV shows that the mixing parameter, as well as the
inelasticities, are very small~\cite{ARN2007}. To a good approximation, we
may therefore neglect the coupling and use the Watson theorem also for the
individual $^{3\!}P_2$ and $^{3\!}F_2$ partial waves, where the phases are
$\delta_{^{3\!}P_2}=17.9^{\circ}$ and $\delta_{^{3\!}F_2}\approx
0^{\circ}$~\cite{ARN2007}.

Two potential models also suggest that the $^{3\!}P_2-\,^{3\!}F_2$ channel
coupling is weak~\cite{MAC1987,HAI1993}. The quality of this approximation
was also checked by explicit calculations of the $d$-wave production
amplitudes within chiral effective field theory up to order $m_{\pi}/m_N$
(NNLO)~\cite{BAR2013}. These show that the phase assumptions made here should
be valid to within $\pm 2^{\circ}$. It should, however, be noted that we do
not neglect the channel coupling in the $^{3\!}S_1-\,^{3\!}D_1$ case, where
the associated effects can be very strong. The phases of the $I=0$ amplitudes
$M_p^S$ and $M_p^D$ are determined in the fits through their interferences with
the $I=1$ amplitudes.

\section{Experiment}
\label{sec3}

\begin{figure}[htb]
\centering
\includegraphics[width=\columnwidth]{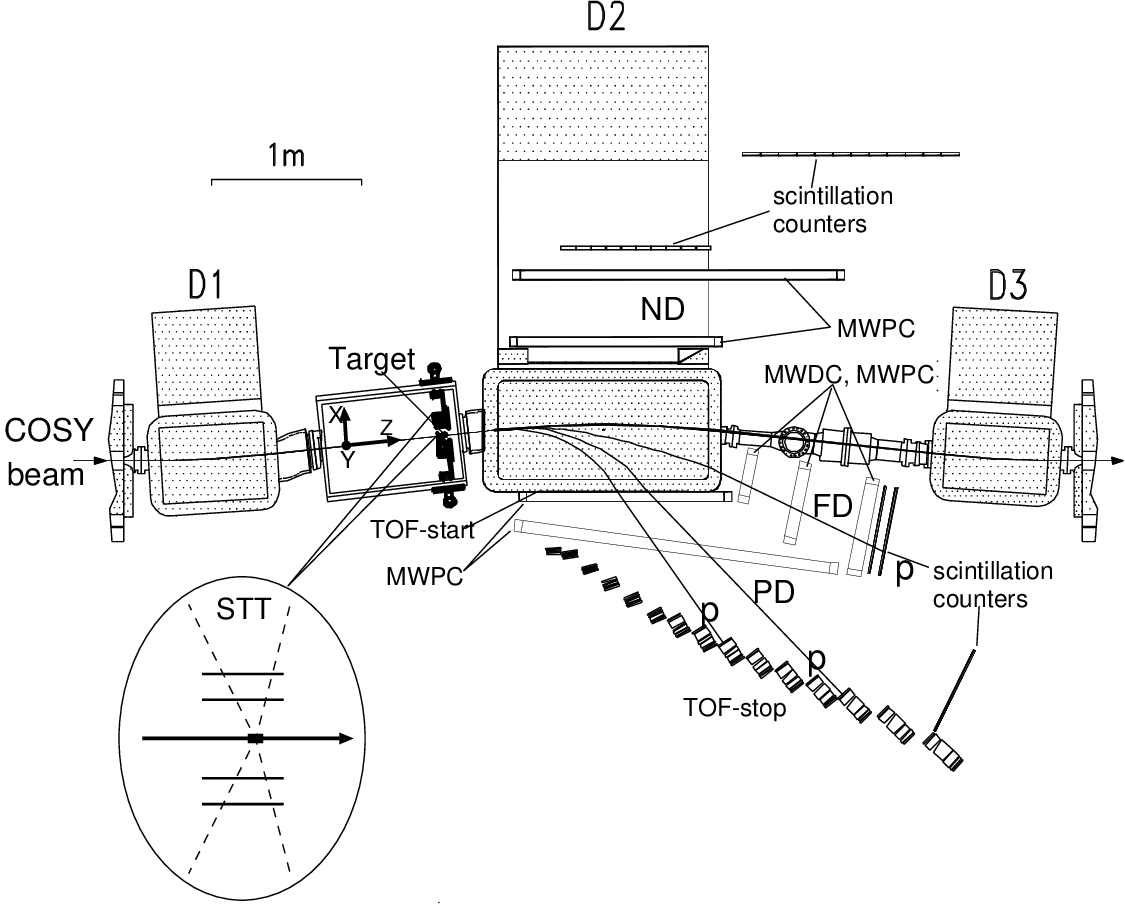}
\caption{Top view of the ANKE spectrometer setup, showing the positions of
the Positive (PD), Negative (ND), and Forward (FD) detectors, as well as the
Silicon Tracking Telescope (STT). The dipoles D1 and D3 deflect the
circulating proton beam in and out of the spectrometer, whereas D2 serves
as an analyzing
magnet. Typical proton trajectories (labeled $p$) in the FD and PD systems
are indicated, as are the axes of the coordinate system at the target.}
\label{fig:ANKE}
\end{figure}

The experiments were carried out with the ANKE magnetic spectrometer installed at
an internal beam position of the Cooler Synchrotron (COSY) at the
Forschungszentrum J\"ulich~\cite{BAR1997}. The ANKE magnetic system comprises
three dipole magnets. The spectrometric magnet D2 is used for the momentum
analysis of the reaction products while D1 and D3 deflect the circulating
beam onto the target and back to the nominal orbit, respectively. The
spectrometer contains several groups of detectors that are shown in
Fig.~\ref{fig:ANKE}. The positive side (PD)  and
forward  (FD)~\cite{CHI2002,DYM2003} detectors are used for the fast positively charged
ejectiles and the negative side detector (ND) for the negatively charged
ones. Each of these groups contains a set of multiwire proportional or drift
chambers for track reconstruction and scintillation counters for triggering
and measuring the arrival time of the particle.

The program for studying the $NN\to NN\pi$ pion production was started by
making measurements of the differential cross section and vector analyzing
power with a polarized proton beam incident on unpolarized cluster-jet
$\rm{H}_2$ and $\rm{D}_2$ targets~\cite{KHO1999}. The deuterium target
allowed the \pnpppi\ reaction to be investigated in quasi-free kinematics in
proton-deuteron collisions. The \pppi\ and \pnpppi\ measurements were carried
out one after the other, with the same settings of the beam and detectors.
This reduced possible systematic uncertainties in the subsequent combined
analysis of the data.

To select quasi-free kinematic conditions, the momentum of the spectator
proton has to be reconstructed for each event. In the proton-deuteron
experiment, some of these protons were detected in a Silicon Tracking
Telescope (STT)~\cite{SCH2003} installed in the ANKE target chamber close to
the interaction region, as shown in Fig.~\ref{fig:ANKE}. Since no time
information from the STT was used in the original work~\cite{DYM2012}, the
main source of background in this analysis was from accidental coincidences
between a fast proton pair and a slow spectator proton produced in a separate
interaction. In Sec.~\ref{sec:pdAy} of the present paper we give the results
of a refined analysis of these data that exploits the time information, thus
reducing dramatically the accidental background.

The double-polarized measurements reported here were carried out in inverse
kinematics, with the polarized deuteron beam incident on the ANKE polarized
target~\cite{MIK2013}. This approach has the advantage that the acceptance of
the spectator protons, emitted in the forward direction at about half the
beam momentum and detected in the ANKE forward detector
(Fig.~\ref{fig:ANKE}), is significantly higher than in the STT case. This is
largely due to the absence of a lower cut on the energy of the
spectator proton. It must however also be remarked that the ANKE deuterium
polarized target had not been commissioned by the time of the current
experiment, whereas the hydrogen one had already been tested and successfully
used in other experiments. The final diproton pair from the \nppppi\ reaction
was recorded in the PD.

Measurements with a polarized deuteron beam and polarized hydrogen target at
ANKE have already been described in some detail~\cite{MCH2013}. In our case,
only the vector polarization modes of the beam, with ideal values of
$P_{\uparrow}=\frac{2}{3}$ and $P_{\downarrow}=-\frac{2}{3}$, were used but
the actual values depend on the precise adjustments of the hyperfine
transition units in the source. The polarizations measured at the beginning
and end of the experiment at the injection energy of 75.6~MeV with the COSY
low energy polarimeter differed slightly in magnitude from each other, with
$P_{\uparrow}=+61\pm 4$\% and $P_{\downarrow}=-50\pm 3$\% for the two modes.
The tensor polarization for both modes was estimated to be below 2\%. No
depolarizing resonances exist for deuterons in the COSY energy range and so
these polarizations should be preserved after acceleration in COSY. However,
we cannot guarantee that these results hold throughout the experiment.
Accurate polarimetry can only be carried out by using the double-polarization
data themselves.

The injected beam was electron cooled and a stacking procedure applied to
increase the beam intensity~\cite{STE}. Typically, ten stacks were made with 30 seconds
cooling time for each, resulting in $3-12\times 10^{9}$ deuterons being stored
and accelerated. The time loss associated with the stacking was low compared
to the total cycle length of 30 minutes. The beam was polarized
perpendicularly to the machine plane, and the spin direction reversed every
cycle.

In order to increase the target density, the gas from the atomic beam source
(ABS)~\cite{MIK2013} was fed into a Teflon-coated ($25~\mu$m thick) aluminum
storage cell with dimensions $x\times y\times z=15\times 19\times
390$~mm$^3$. This resulted in a density of $\sim10^{13} \rm{atoms/cm^2}$ and
luminosities of up to $\sim 10^{29} \rm{cm}^{-2}\rm{s}^{-1}$. The ABS
produced a jet of atomic hydrogen that was polarized perpendicularly to the
COSY beam direction, and the spin orientation of the atoms in the cell was
aligned with the vertical field of the D2 dipole. Proton polarizations of
above 90\% were achieved with the jet~\cite{MIK2013}. The direction of the
spin was reversed every five seconds.

Unlike the measurements with the cluster-jet target, the main source of
background with the polarized target was the interactions of the beam
particles with the aluminum cell walls. In these interactions, the same
processes occur on the nucleons in the nuclei in the cell walls as in the
hydrogen. To study the properties of the background, dedicated measurements
were conducted with both an empty cell and with nitrogen gas in the cell.

The polarizations of the nucleons in the beam and target were studied using
the data on the \vnpdpi\ reaction taken in the $\pol{d}p \to p_{\text{spec}}d
\pi^0$ process that were recorded simultaneously with those of \nppppi. In
both processes, the spectator proton was detected in the FD
(Fig.~\ref{fig:ANKE}), and this detector was also used for the deuteron
formed in the \npdpi\ reaction.

In order to study the  beam and target polarizations for each spin direction,
some data were also taken with an unpolarized beam, an unpolarized target,
and both of them unpolarized. For these purposes, the COSY unpolarized beam
source was used or the cell was filled from the source of unpolarized
$\rm{H}_2$ gas.
%
%

\section{Data analysis in the deuteron-proton experiment}
\label{sec4}
\subsection{Reaction selection}

The identification of the \nppppi\ reaction starts with the selection of the
final diprotons via the time-of-flight (TOF) criterion, as described in
Ref.~\cite{MCH2013}. The difference of the arrival times measured in the
scintillation counters of the FD and PD is compared to that calculated under
the assumption that the two particles are protons. The spectator proton
detected in the FD is then identified by the TOF difference built for it and
each of the protons in the diproton. The selection of the final $dp$ pairs
from the $dp\to d\pi^0p_{\text{spec}}$ reaction was done analogously. The
protons detected in the PD can also be selected using the TOF between the
start and stop scintillation counters, shown as TOF-start and TOF-stop in
Fig.~\ref{fig:ANKE}.

The $^{1\!}S_0$ events are selected from among the pairs of identified
protons by applying a cut on the excitation energy in the pair
$E_{pp}<3$~MeV. This can be done reliably due to the excellent resolution of
$\sigma({E}_{pp})<0.3$~MeV in this ${E}_{pp}$ region.

\begin{figure}[htb]
\centering
\includegraphics[width=\columnwidth]{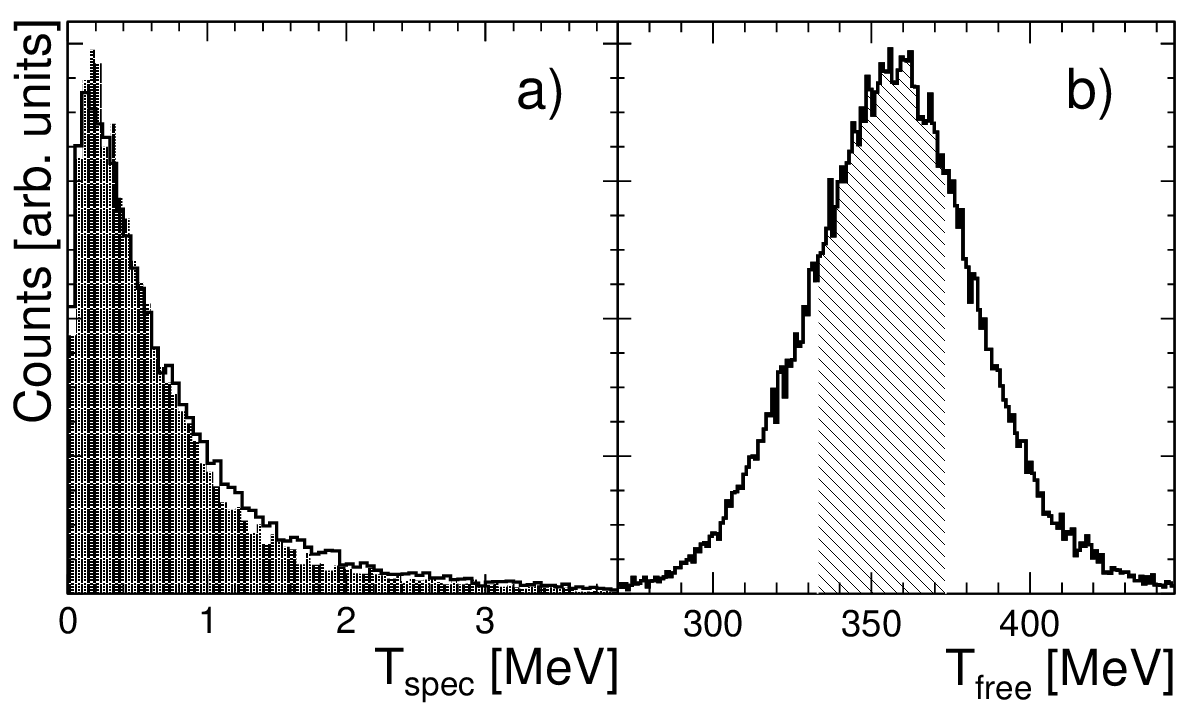}
\caption{a) Experimental (histogram) and simulated (shaded area) distribution
in the deuteron rest frame of spectator proton energies from the $dp\to
\{pp\}_{\!s}\pi^-p_{\text{spec}}$ reaction for the selected $T_{\text{free}}$
range. b) Experimental distribution of effective beam energies; the region
selected for analysis is shown by the shading.} \label{fig:TspTfree}
\end{figure}

It can be seen from the experimental distributions of spectator proton
energies $T_{\text{spec}}$ in the deuteron rest frame shown in
Fig.~\ref{fig:TspTfree}a that the acceptance is particularly favorable for
small $T_{\text{spec}}$. The quasi-free kinematical conditions are ensured by
choosing energies $T_{\text{spec}}<6$~MeV. The value of the spectator proton
3-momentum allows one to evaluate the effective beam energy $T_{\text{free}}$
in the \nppppi\ reaction,
\begin{equation}
T_{\text{free}}=(s-(M_p+M_n)^2)/2M_p,
\end{equation}
where $\sqrt{s}$ is the total cm energy in the $np$ system and $M_p$ and
$M_n$ are the proton and neutron masses. Figure~\ref{fig:TspTfree}b shows the
experimental distribution of $T_{\rm{free}}$ and the region of
$T_{\text{free}}=(353\pm 20)$~MeV that is retained in the analysis.

After application of the selections described above, the \nppppi\ and \npdpi\
reactions could be identified by comparing the missing-mass peaks in the data
with the masses of the missing pions.

\subsection{Background subtraction}

The main source of background consisted of events corresponding to the
reactions studied, but produced on the nucleons of the aluminum in the cell
walls. The only difference between good $dp$ data and background was in the
shape of the missing-mass distributions. Due to the Fermi smearing in the
aluminum, these were significantly wider for the background. Dedicated
measurements with the empty cell and filled with nitrogen (N$_2$) gas were
carried out to derive the shape of the background. Although more accurate
background distributions could be obtained through measurements with the
empty cell, it was realized that collecting the necessary statistics would
require too much time. The N$_2$ target, which produced a signal similar to
that from aluminum, led to substantially higher statistics.

\begin{figure}[htb]
\centering
\includegraphics[width=\columnwidth]{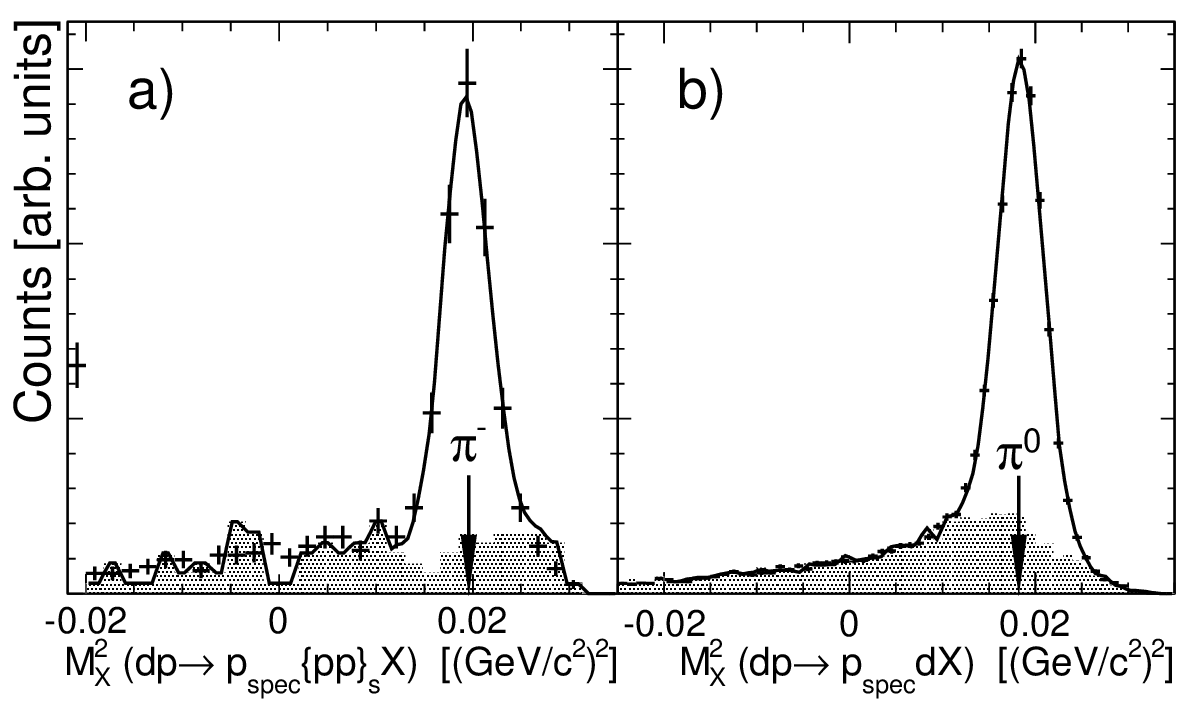}
\caption{ Missing-mass-squared for the a) $dp\to
p_{\text{spec}}\{pp\}_{\!s}X$ and b) $dp\to dpX$ reactions. The data with
error bars were obtained with hydrogen in the cell. The shaded area shows the
scaled background and the line the total fitted function.}
 \label{fig:bkgCell}
\end{figure}

Figure~\ref{fig:bkgCell} shows the missing-mass spectra for the \nppppi\ and
\npdpi\ reactions obtained with the hydrogen cell target. Each spectrum was
fitted by the sum of a Gaussian and a scaled distribution of the background
collected with N$_2$ in the cell. The shape of the background was determined
separately for all the bins in the polar and azimuthal angles used in the
analysis.

\subsection{Reconstruction of the kinematics}

The reconstruction of the momentum of a particle passing through the
analyzing magnet D2 relies on the information on the position of the
interaction vertex. In the case of a storage cell, this is known much less
precisely than for the point-like cluster-jet target. Uncertainties in the
interaction vertex location lead to poorer resolution for the momentum of the
detected particle, as well as to systematic deviations in its value. To
alleviate this problem, a vertex reconstruction procedure was applied that
fits simultaneously the trajectories of the particles and their arrival
times. The parameters of the fit are the three-momenta of the particles and
the longitudinal coordinate $Z$ of the vertex. The accuracy achieved in the
coordinate was $\sigma(Z)= 5.6$~cm for the $dp\to pp\pi^-p_{\text{spec}}$
process and $\sigma(Z)= 12$~cm for $dp\to d\pi^0p_{\text{spec}}$. Knowing $Z$
even with such a limited precision allows one to reject interactions with the
unpolarized gas in the target vacuum chamber outside of the cell. The
resulting polar and azimuthal angular resolutions achieved for both the
\nppppi\ and \npdpi\ reactions were in the $\sigma(\vartheta_\pi)=2-6^\circ$
and $\sigma(\varphi_\pi)=4-10^\circ$ ranges, depending on the angular
regions. The resolution in spectator energy was better than 0.15~MeV and that
in effective beam energy about 4~MeV.

\subsection{Relative normalization. Polarimetry}
\label{sec:relLum}

In order to extract values of the polarization observables, one has to know
the ratios of the integrated luminosities for each of the beam and target
spin orientations. This could be done straightforwardly if the experiment
involved only single-polarized data by comparing the numbers of counts in
kinematical regions where the analyzing power vanishes such as, for example,
in the forward direction for the \npdpi\ reaction. Using the vector analyzing
power available at 353~MeV for the $\pol{p}p\to d\pi^+$ reaction from the
SAID database~\cite{ARN1993}, one could then determine either the beam or
target polarization. The asymmetry of counts for each spin direction with
respect to the properly normalized unpolarized counts yields the
polarizations for each direction separately. The results obtained from a
limited sample of single-polarized data for the target (proton) 
polarization were
$Q_\uparrow=59\pm 7\%$ and $Q_\downarrow=-70\pm11\%$. Those for the beam (neutron) were
$P_\uparrow=55\pm 8\%$ and $P_\downarrow=-45\pm 8\%$.  Within the
large error bars the two sets of magnitudes are consistent and only a very
small error is introduced by assuming this equality in the subsequent
analysis.

\begin{figure}[htb]
\centering
\includegraphics[width=\columnwidth]{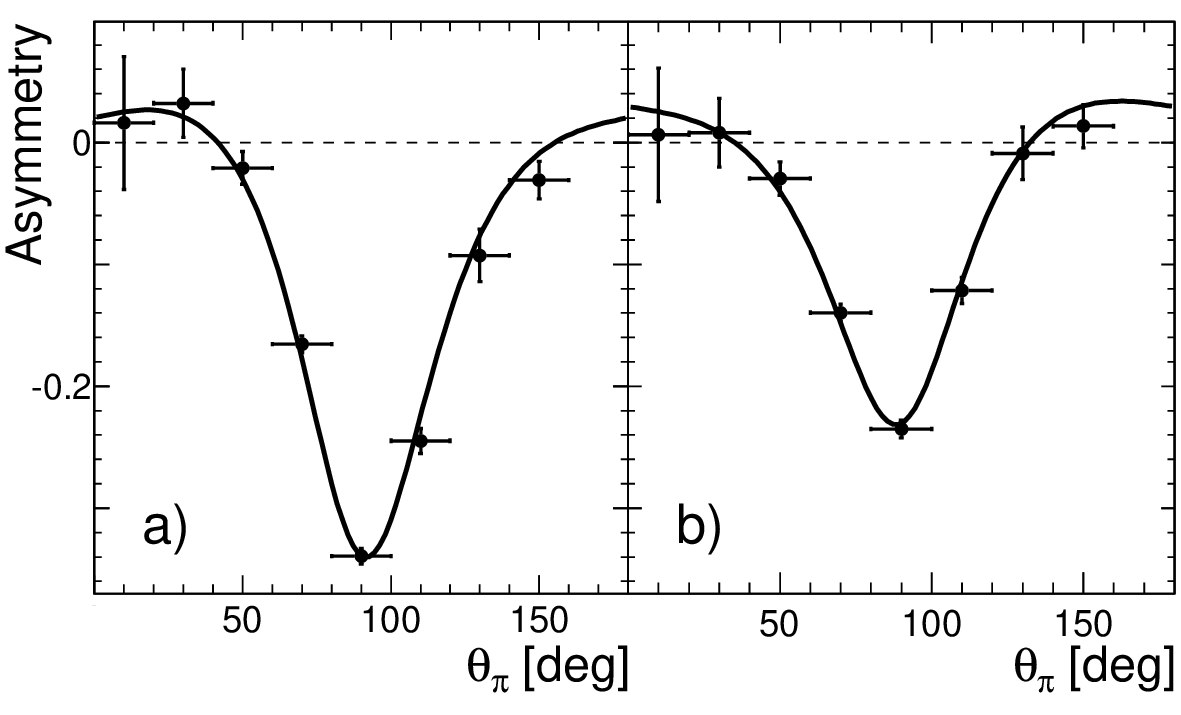}
\caption{Experimental asymmetry in the quasi-free \vnpdpi\ data for (a) the
target polarization and (b) the beam polarization. The curves show the SAID
predictions for $A_y$ in $\pol{p}p\to d\pi^+$ at 353~MeV~\cite{ARN1993},
scaled with the value of polarization to fit the data, which are shown with
their statistical errors.}
 \label{fig:pol}
\end{figure}

It is perhaps comforting that the neutron beam polarizations obtained here are
both consistent with a factor of 0.9 times the deuteron vector polarizations
measured at injection at the beginning and end of the experiment.
Furthermore, the differences in the magnitudes of the polarizations with spin
up or down are relatively small. However both the beam and target
polarization results given here are merely indicative; it is absolutely
necessary to deduce their averages over the course of the experiment and this
means extracting the averages from the double-polarized data. This also has
the advantage that much larger statistics are then available.

As can be seen from Eq.~(\ref{asymmetry}), in this double-polarized case the
polarization effects in the $\pol{n}\pol{p}\to d\pi^0$ cross section do not
disappear in the forward direction and, to exploit such data for polarimetry
purposes, one has to know the spin-correlation coefficients for the reaction
as well as the analyzing power. Under our experimental conditions, in a first
approximation, the integrated luminosities can be assumed to be equal so that
one can sum the beam polarization modes to define polarization of the target
and vice versa. Using the transverse spin-correlation coefficients for
$\pol{p}\pol{p}\to d\pi^+$ at 353~MeV~\cite{ARN1993}, values of the
polarizations could be deduced and these led to the following estimates for
the luminosity ratios:
$R_{\uparrow\downarrow}={L_{\uparrow\downarrow}}/{L_{\uparrow\uparrow}}
=0.97\pm 0.02$,
$R_{\downarrow\uparrow}={L_{\downarrow\uparrow}}/{L_{\uparrow\uparrow}}
=0.95\pm 0.02$ and
$R_{\downarrow\downarrow}={L_{\downarrow\downarrow}}/{L_{\uparrow\uparrow}}
=1.05\pm 0.02$, $R_{\uparrow\uparrow}\equiv 1$.

Although the luminosity ratios are, as expected, close to unity, the
deviations from this are important in the extraction of the polarizations.
Inserting these into the analysis, the mean magnitudes of the beam and target
polarizations were found to be $|P|=50\%\pm 3\% (\text{stat}) \pm 3.5\%
(\text{syst})$ and $|Q|=69\%\pm 2\% (\text{stat}) \pm 3.5\% (\text{syst})$.
The 3.5\% systematic errors arise from uncertainties in the
$\pol{p}\,\pol{p}\to d\pi^+$ calibration reaction~\cite{ARN1993}. These
results are illustrated in Fig.~\ref{fig:pol}, where the experimental
asymmetries in the $\pol{n}p\to d\pi^0$ counts are fitted by the scaled
analyzing power for the $\pol{p}p\to d\pi^+$ reaction. We should stress that
here and elsewhere in the paper we take $\theta_{\pi}$ to be the c.m.\ angle
of the pion with respect to the incident proton direction, as we did for our
deuterium target data~\cite{DYM2012}.

It should be noted that, since the spin-correlation parameter $A_{y,y}$ for
the \vpnpppi\ reaction is constrained to be unity, the product $PQ$ of the
polarizations can be independently determined from an analysis of the
reaction data themselves.

\section{Results}
\label{sec5}

\subsection{Reanalysis of the proton beam data}
\label{sec:pdAy}

In Ref.~\cite{DYM2012} we presented results from measurements of the
unpolarized differential cross section and proton analyzing power of the
$\pol{p}n\to \{pp\}_{\!s\,}\pi^-$ reaction with the polarized proton beam and
a deuterium cluster-jet target. The spectator protons were here detected in
one of the Silicon Tracking Telescopes that were installed in the ANKE target
chamber (Fig.~\ref{fig:ANKE}). Each STT, located to the left and right of the
beam, contained two layers of silicon strip detector. Only protons passing
through the first layer and stopping in the second were analyzed, and this
limited the energy of the slow spectator proton to be in the range
 $2.6 <T_{\text{spec}}<6$~MeV. The energy loss in the two layers allowed one to separate protons
from deuterons~\cite{RIE1980}, while the two-dimensional coordinates measured
in each layer enabled the reconstruction of the direction vector and the
coordinates of the interaction point. The latter was used to help in the
background suppression.

\begin{figure}[htb]
\centering
\includegraphics[width=\columnwidth]{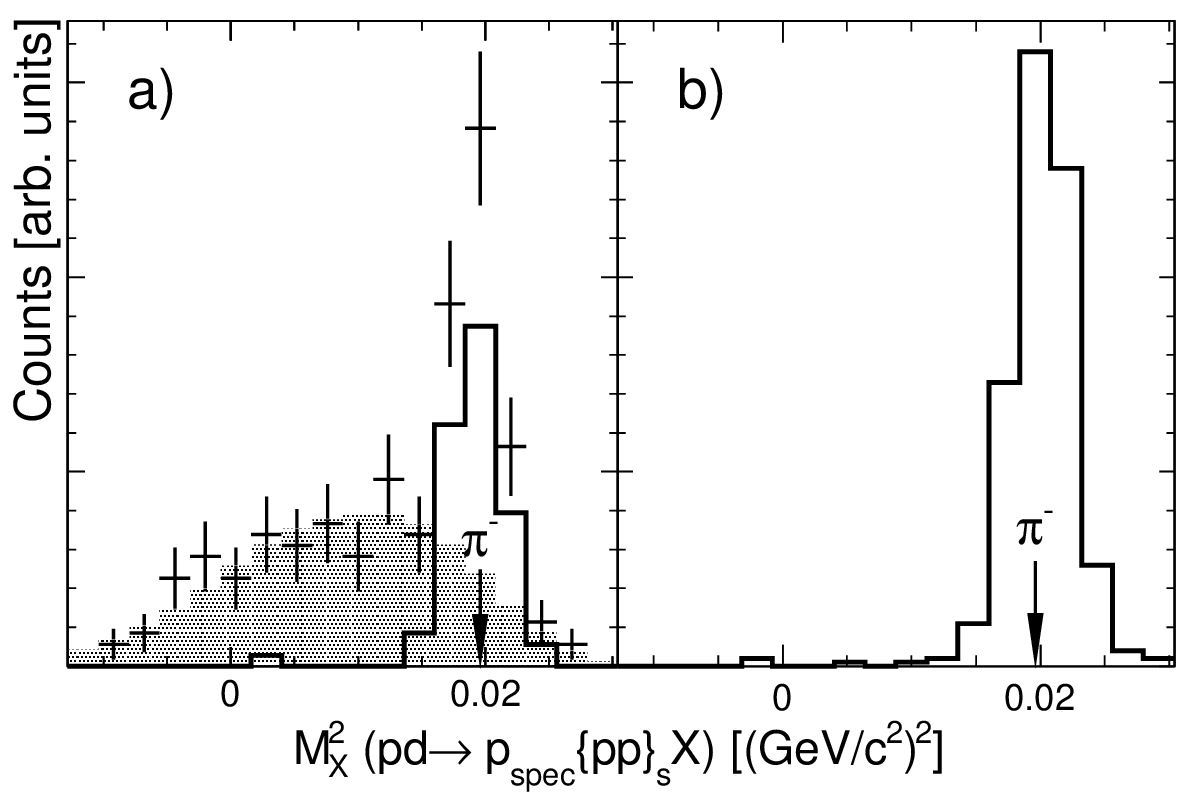}
\caption{The experimental $pd\to p_{\text{spec}}\{pp\}_sX$
missing-mass-squared spectrum. a) Spectrum measured using the first two STT
layers without the time criterion (points with error bars) and the
constructed and scaled background (filled area)~\cite{DYM2012}. The data
obtained using the time criterion (line) show almost no background. b)
Spectrum for the same angular bin and detector combination obtained from the
first STT layer only by using the timing information.}
 \label{fig:timing}
\end{figure}

The major source of background in this approach was accidental coincidences
with a random spectator proton in the STT. The background level in the pion
peak of the \pnpppi\ data was up to 50\% and a special procedure had to be
derived in order to evaluate its shape in the missing-mass
spectra~\cite{DYM2012}. A way to access the time information from the STT has
recently been established and this has allowed the deuterium target data to
be reanalyzed with a substantially lower background, as demonstrated by the
results shown in Fig.~\ref{fig:timing}a.

Suppressing the accidental background also allowed one to increase the
statistics by considering also the slower spectator protons that stopped in
the first layer of the STT. This was possible because, at a beam energy of
353~MeV, the $pd\to ppp\pi^-$ reaction is the only process that can result in
three positively charged hadrons in the final state. One can therefore
identify the reaction, merely by making a missing-mass selection, without
explicitly identifying the spectator proton through its energy loss. The
second layer of the STT served as a veto for such events. The center of the
beam-target interaction region was used as the starting point on the track
that defined the direction of the momentum vector. Although the accuracy of
the measurement of the spectator three-momentum was poorer in this case, it
was still quite sufficient for the identification of the \pnpppi\ reaction by
calculating the mass of the missing pion. Without exploiting the time
information, the level of background for such events would made this sample
practically unusable. After application of this criterion the background was
reduced to the few percent level, as shown in Fig.~\ref{fig:timing}b.
However, since the overall acceptance is much harder to estimate for these
events, they were only used to improve the measurement of the analyzing
power. This approach increased the statistics for $\theta_{\pi}>30^{\circ}$
by about a factor of two. The improvement was far less significant at smaller
angles, where the pions were directly measured in the negative side detector.

\begin{figure}[htb]
\centering
\includegraphics[width=\columnwidth]{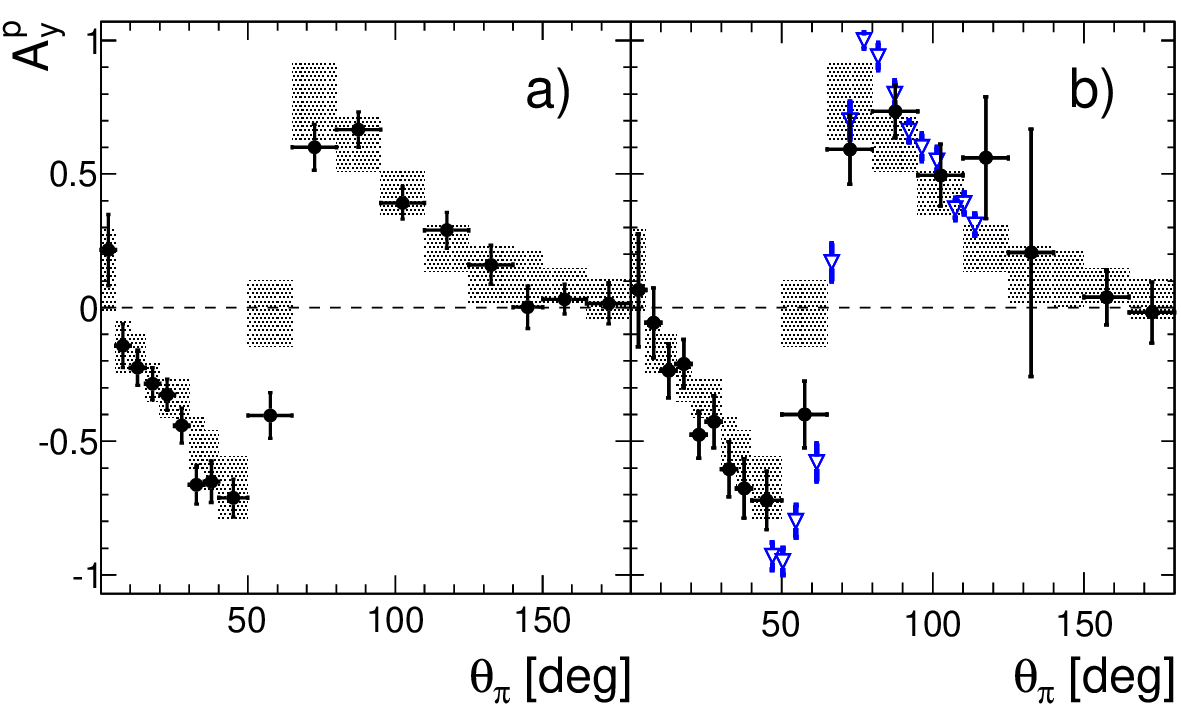}
\caption{(Color online) Analyzing power $A_y^p$ for the quasi-free
$\pol{p}n\to \{pp\}_{\!s}\pi^-$ reaction at 353~MeV with the ${E}_{pp}<3$~MeV
cut. The published ANKE data~\cite{DYM2012} are shown with the shaded error
bands. The black circles show the reanalyzed ANKE data with statistical
errors using ${E}_{pp}$ cuts of a) 3~MeV and b) 1.5~MeV. The TRIUMF
data~\cite{HAH1999} are represented by the blue triangles.}
\label{fig:oldNewAy}
\end{figure}

The enhanced background suppression and improved spectator proton energy
range led also to a better determination of the beam polarization. The
relative polarization uncertainty of 11\% introduced the largest systematic
error in the combined amplitude analysis of the \pppi\ and \pnpppi\
reactions~\cite{DYM2012}. The new value of the beam polarization of $63.3\pm
3.6$ \%, where the error is dominated by that of the calibration
data~\cite{ARN1993}, reduces this uncertainty by a factor two.

The increased statistics allowed us to test the effect of imposing the
tighter limitation on the diproton excitation energy, ${E}_{pp}<1.5$~MeV,
that was used at TRIUMF~\cite{HAH1999}. Figure~\ref{fig:oldNewAy} shows the
newly obtained $A_y^p$ data for the two ${E}_{pp}$ cuts compared with the
published ANKE and the TRIUMF data. It is seen that the analyzing powers are
little changed through the introduction of the harder $E_{pp}$ cut that was used
for the TRIUMF data~\cite{HAH1999}.

\subsection{Extraction of the unpolarized cross section and analyzing power
from double-polarized data}

The study of the \vnvpppppi\ reaction in $dp$ kinematics, and with a
different set of detectors, opens the possibility of checking the systematic
uncertainties in the published ANKE data on $d\sigma/d\Omega$ and
$A_y^p$~\cite{DYM2012}, as well as the consistency of the more complicated
analysis of the double polarization measurements with the long cell target.
The overall statistics collected in the two experiments are similar but, due
to the difference in acceptance, the angular distributions of raw events
differ significantly.

\begin{figure}[htb]
\centering
\includegraphics[width=\columnwidth]{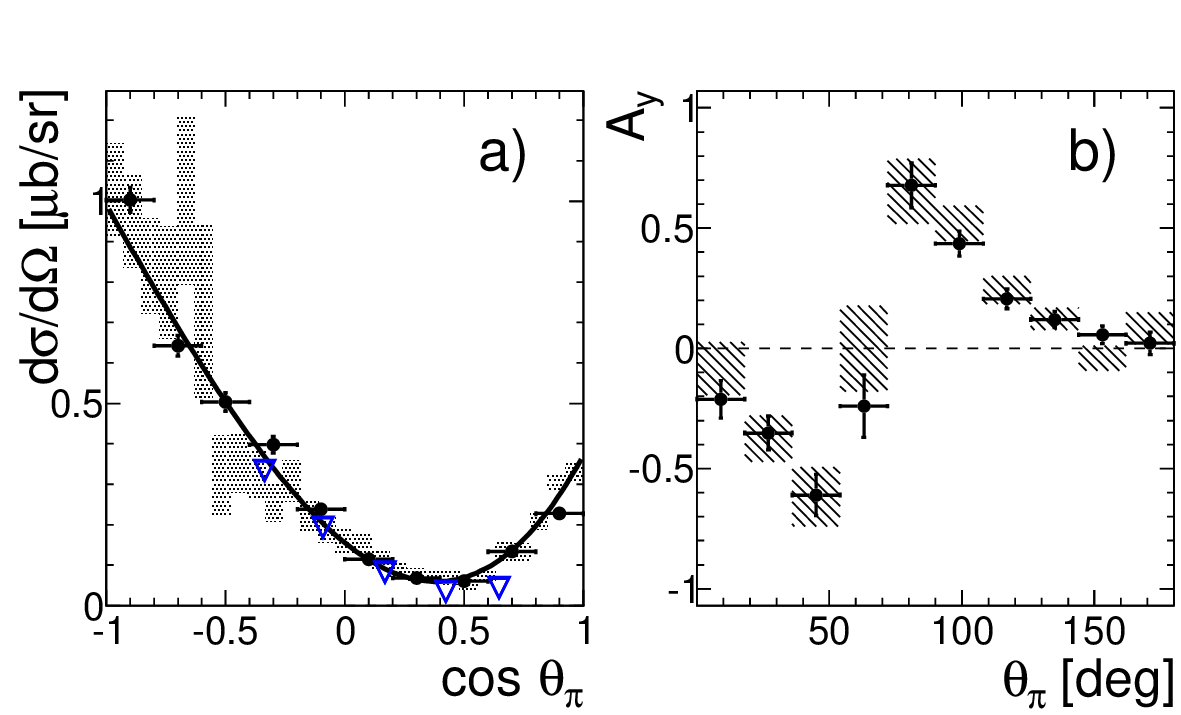}
\caption{(Color online) Observables measured for the \pnpppi\ reaction at
353~MeV using the ${E}_{pp}<3$~MeV cut. a) The unpolarized differential cross
section. The published ANKE data~\cite{DYM2012} are shown by the shaded error
bands and the curve is a direct cubic fit to these data. The black circles
show the data with statistical errors obtained in the new ANKE
double-polarized experiment. Since no absolute normalization was here
achieved, these values were scaled by an arbitrary overall factor. The TRIUMF
data~\cite{DUN1998} are shown by the blue triangles. b) The proton analyzing
power $A_y^p$ (points) and neutron analyzing power $A_y^n$ (shaded bands)
obtained simultaneously in the double-polarized experiment.}
 \label{fig:newXsecAy}
\end{figure}

It is much harder to evaluate a precise value for the absolute normalization
for data taken with a long cell target than it is for a point-like vertex
experiment. We therefore present in Fig.~\ref{fig:newXsecAy}a only an
arbitrarily scaled cross section obtained from the cell data to facilitate a
comparison of its shape with the published results from ANKE~\cite{DYM2012}
and TRIUMF~\cite{DUN1998}. There is reasonable consistency between the two
ANKE experiments, in particular in the forward pion angle region where the
TRIUMF data begin to deviate.

The proton and neutron analyzing powers were extracted from the
double-polarized data by averaging over the beam and target polarization
states, respectively. Although the polarization of the beam neutron is less 
than that of the target proton, and so the error bars are larger, the values 
of $A_y^p$ and $A_y^n$ shown in Fig.~\ref{fig:newXsecAy}b are mutually 
compatible. This provides extra evidence to support the validity of the 
current experiment and its analysis. The new results, which are completely 
consistent with the published ANKE data, have comparable statistical and 
systematic uncertainties and so can be used in a combined analysis.

\subsection{Measurement of the spin-correlation coefficients
$\boldsymbol{A_{x,x}}$ and $\boldsymbol{A_{y,y}}$}

It follows from Eq.~(\ref{asymmetry}) that the experimental double-polarized
asymmetry $\xi$ can be written as:
\begin{equation}
\xi  = \frac{\Sigma_1-\Sigma_2}{\Sigma_2+\Sigma_2}, \;\;\;\;\;
\frac{\xi}{PQ} = A_{x,x}\sin^2\varphi_{\pi}+A_{y,y}\cos^2\varphi_{\pi},
\label{eq:expAsym}
\end{equation}
where $\Sigma_1=N_{\uparrow\uparrow}+N_{\downarrow\downarrow}$ and
$\Sigma_2=N_{\uparrow\downarrow}+N_{\downarrow\uparrow}$. Here $N$ represents
the number of events collected with the directions of the beam and target
spins indicated by the arrows, normalized to the corresponding relative
luminosity $R$, defined in section~\ref{sec:relLum}. $PQ$ is the product of
the beam and target polarizations.

The background was subtracted separately from the combinations
$\Sigma_1-\Sigma_2$ and $\Sigma_1+\Sigma_2$. Since the background
contribution showed practically no polarization dependence, its effects were
very low in the difference.

\begin{figure}[htb]
\centering
\includegraphics[width=\columnwidth]{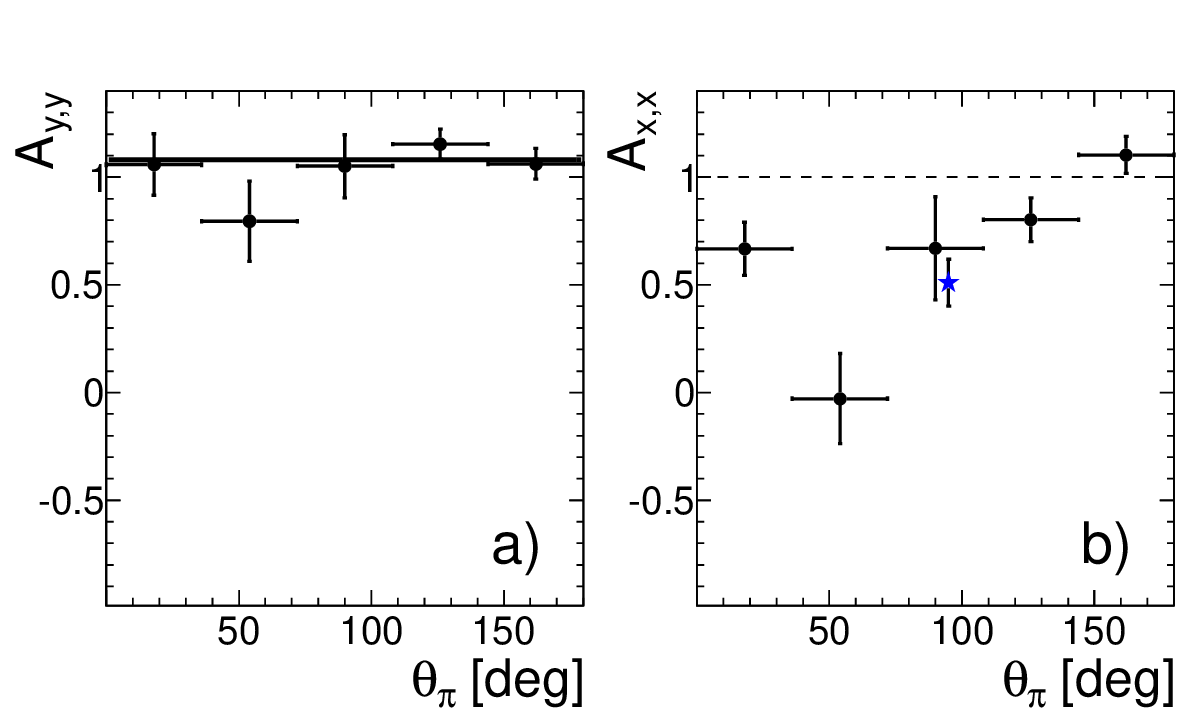}
\caption{(Color online) Spin-correlation coefficients for the \vnvpppppi\
reaction at 353~MeV with statistical errors as a function of the pion
emission angle. a) $A_{y,y}$. The horizontal line represents a fit with a
constant value. b) The values of $A_{x,x}$ were subsequently deduced by
demanding that $A_{y,y}=1$ for all pion angles. The systematic uncertainties
of the  $90^{\circ}$ point (blue star, shifted for visibility) deduced  from
Eq.~(\ref{Axx_rel}) are much larger than the purely statistical errors shown.
\label{fig:AxxAyy}}
\end{figure}

The experimental data were divided into five bins in the pion emission angle
$\theta_{\pi}$ and $\xi/PQ$ was fitted as a linear function of
$\cos^2\varphi_{\pi}$ in each bin. The acceptance of the apparatus was
significantly higher for events with large $\cos^2\varphi_{\pi}$ so that the
value of $A_{y,y}$ was determined with smaller uncertainty than that of
$A_{x,x}$. The results for both spin-correlation parameters are shown in
Fig.~\ref{fig:AxxAyy} as a function of $\theta_{\pi}$.

As can be seen from Eq.~(\ref{eq:expAsym}), the experiment is essentially
self-analyzing. The values obtained for $A_{x,x}$ and $A_{y,y}$ depend only
on the product of the beam and target polarizations $PQ$ and this can be
determined by requiring that $A_{y,y}=1$ when averaged over all pion angles.
This property of the experiment also provides a powerful tool to study
systematic uncertainties in the measurement. The fitted constant value of
$A_{y,y}=1.08\pm 0.04$ is consistent with unity when one takes into account
the 0.11 uncertainty coming from the error in the polarization product. By
demanding that $A_{y,y}=1$, one obtains the more precise measurement of the
polarization product, $PQ=0.373\pm 0.015$, that can then be used in the
determination of $A_{x,x}$.

In order to reduce the uncertainty in the extraction of $A_{x,x}$, it was
assumed that $A_{y,y}=1$ for all $\theta_{\pi}$ and the $\cos^2\varphi_{\pi}$
fit repeated. It was this procedure that led to the results shown in
Fig.~\ref{fig:AxxAyy}b. The uncertainty in $A_{x,x}$ is dominated by
statistics. The systematic uncertainties originated from (\textit{i}) the
polarization product error (0.04), (\textit{ii}) the possible difference in
the up/down polarizations (0.01), (\textit{iii}) the relative normalization
uncertainty (0.023), and (\textit{iv}) the effect of a longitudinal spin
component arising from the Fermi motion in the deuteron (up to 0.07).

Also shown in Fig.~\ref{fig:AxxAyy}b is the value at $90^{\circ}$ of
$A_{x,x}=0.51\pm0.11$ that is deduced from Eq.~(\ref{Axx_rel}) by using
direct fits to the $\pi^0$ and $\pi^-$ production cross sections. However,
the error bar quoted here is purely statistical and does not include the
systematic effect from the uncertainty in the relative normalizations.

\section{Partial wave analysis}
\label{sec6}

The partial wave fitting of Ref.~\cite{DYM2012}, described in detail in
Sec.~\ref{sec:pwa}, was repeated taking into account the new data presented
here. These include the reanalyzed $A_y^p$ from the deuterium cluster-jet
experiment, the newly estimated $A_y^p$ from the hydrogen cell data, and the
$A_{x,x}$ results. The values of the unpolarized $(d\sigma/d\Omega)_0$, for
both $\pi^0$ and $\pi^-$ production, and the $A_y^p$ for $\pi^0$ production
were taken from our previous work~\cite{TSI2012,DYM2012}.

In contrast to the procedure adopted in Ref.~\cite{DYM2012}, the squares of
pion $d$-wave amplitudes were not neglected in the analysis. In addition,
effects from the uncertainties in the data normalization were included by
constructing the full non-diagonal covariance matrix $\mathcal{M}$ for the
measured data points and minimizing the general form $\chi^2=\delta_i
\mathcal{M}^{-1}_{ij} \delta_j$, where $\delta_i$ is the $i^{th}$ data-point
residual.

A grid search for $\chi^2$ minima was performed in the space of the
magnitudes and phases of the $p$-wave amplitudes, with the $s$- and $d$-wave
amplitudes being fixed by the fit to the $\pi^0$ data. The search revealed
three minima with very similar values of $\chi^2$. The five amplitudes were
then fitted in the vicinity of each minimum. The properties of the three
solutions are listed in Table~\ref{tab:fit}, where the first minimum
corresponds to that found in Ref.~\cite{DYM2012}. The solutions differ mostly
in the parameters of the $p$-wave amplitudes, while the $s$- and $d$-waves
stay essentially unchanged. The existence of alternative solutions is not
entirely unexpected and some of their properties were studied both
theoretically~\cite{GAL1990} and empirically~\cite{PIA1986}, though neither
of these works considered the case where some of the phases are fixed by the
Watson theorem.

\begin{figure}[htb]
\centering
\includegraphics[width=\columnwidth]{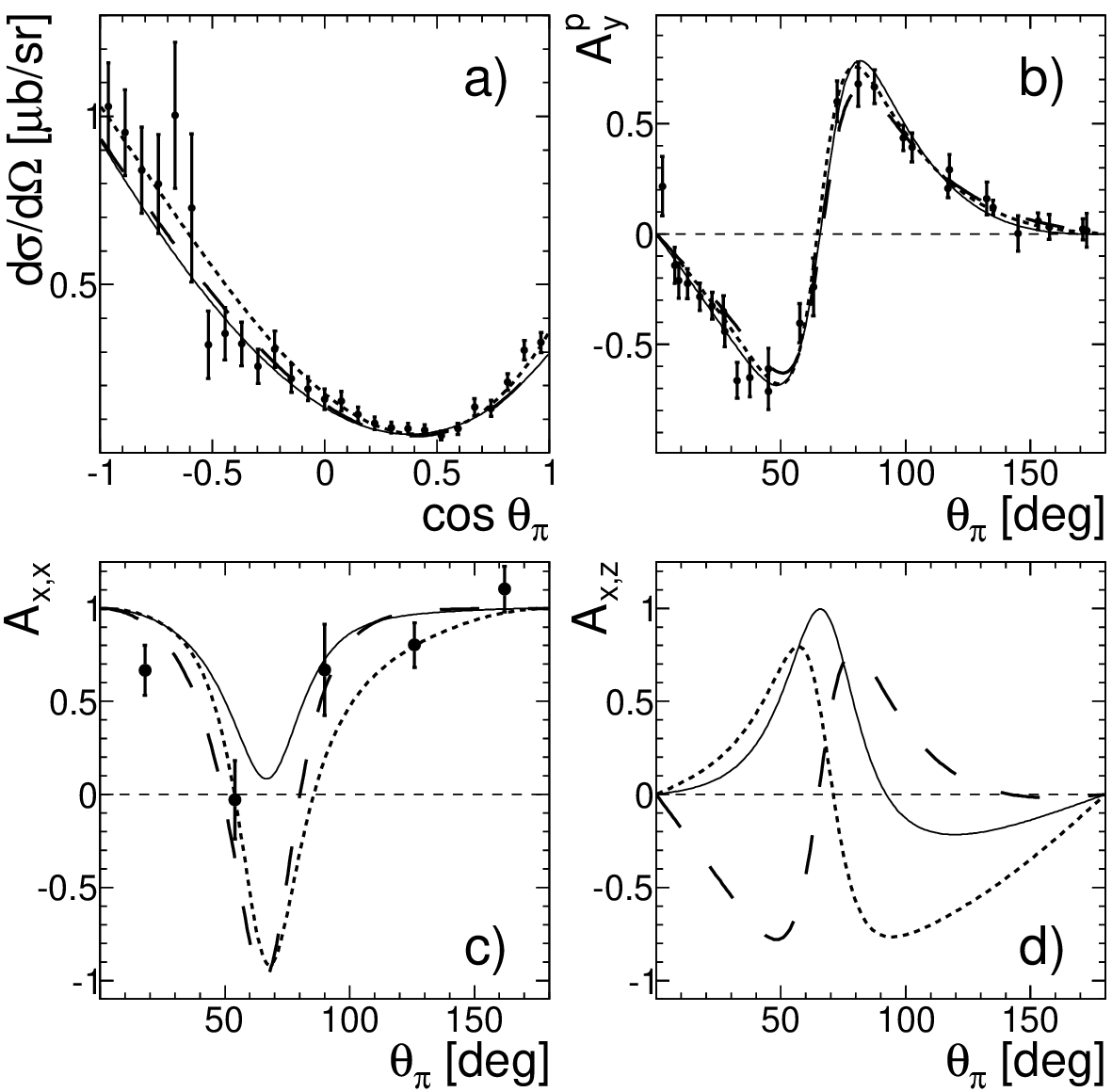}
\caption{Predictions of the partial wave analysis for the \pnpppi\ reaction
at 353~MeV with  the ${E}_{pp}<3$~MeV cut. Also shown are the ANKE
experimental data with statistical errors. The full, long-dashed, and
short-dashed lines correspond to solutions 1, 2, and 3, respectively, as
noted in Table~\ref{tab:fit}. a) Differential cross-section taken from
Ref.~\cite{DYM2012}, b) $A_y^p$ data from this work, c) $A_{x,x}$ data from
this work, d) $A_{x,z}$, for which there are yet no experimental data.}
\label{fig:pwaAll4}
\end{figure}

Figure~\ref{fig:pwaAll4} shows the predictions for the \nppppi\ observables
for the three solutions detailed in Table~\ref{tab:fit} compared with the
ANKE data. The solutions describe equally well the cross section and $A_y^p$
data, but differ significantly in $A_{x,x}$, and especially in $A_{x,z}$. The
$A_{x,x}$ data presented in this work follow the gross features of all three
predictions. However, while favoring somewhat solutions 2 and 3, this is not
decisive given the statistical uncertainties. In view of the drastically
different predictions for $A_{x,z}$, a measurement of this coefficient
becomes especially important for resolving the ambiguities in the analysis
and determining which of the three possible solutions is the physical one.

\begin{table}
\centering
\begin{tabular}{|c|r|r|r|} \hline
Amplitude  &  Real\phantom{xxll} & Imaginary\phantom{l}& Im/Re \phantom{1} \\ \hline
\multicolumn{4}{|c|}{Solution 1: $\chi^2/\textit{ndf}=101/82$ } \\ \hline
$M^P_s$ & $  53.4 \pm    1.0$ & $ -14.1 \pm    0.3$ & \\ \hline
$M^P_d$ & $ -25.9 \pm    1.4$ & $  -8.4 \pm    0.4$ & \\ \hline
$M^F_d$ & $  -1.5 \pm    2.3$ & $   0.0 \pm    0.0$ &\\ \hline
$M^S_p$ & $ -37.5 \pm    1.7$ & $  16.5 \pm    1.9$ &$-0.44\pm0.06$\\ \hline
$M^D_p$ & $ -93.1 \pm    6.5$ & $ 122.7 \pm    4.4$ &$-1.32\pm0.11$ \\ \hline\hline
         \multicolumn{4}{|c|}{Solution 2: $\chi^2/\textit{ndf}=103/82$} \\ \hline
$M^P_s$ & $  52.7 \pm    1.0$ & $ -13.9 \pm    0.3$ &  \\ \hline
$M^P_d$ & $ -28.9 \pm    1.6$ & $  -9.4 \pm    0.5$ &  \\ \hline
$M^F_d$ & $   3.4 \pm    2.6$ & $   0.0 \pm    0.0$ &  \\ \hline
$M^S_p$ & $ -63.7 \pm    2.5$ & $  -1.3 \pm    1.6$ &$0.02\pm0.03$  \\ \hline
$M^D_p$ & $-109.9 \pm    4.2$ & $  52.9 \pm    3.2$ &$-0.48\pm0.03$  \\ \hline \hline
         \multicolumn{4}{|c|}{Solution 3: $\chi^2/\textit{ndf}=106/82$} \\ \hline
$M^P_s$ & $  50.9 \pm    1.1$ & $ -13.4 \pm    0.3$ &  \\ \hline
$M^P_d$ & $ -26.3 \pm    1.5$ & $  -8.5 \pm    0.5$ &  \\ \hline
$M^F_d$ & $   2.0 \pm    2.5$ & $   0.0 \pm    0.0$ &  \\ \hline
$M^S_p$ & $ -25.4 \pm    1.9$ & $  -7.3 \pm    1.5$ &$0.20\pm0.07$  \\ \hline
$M^D_p$ & $-172.2 \pm    5.6$ & $  92.0 \pm    6.2$ &$-0.53\pm0.04$  \\ \hline
\end{tabular}
\caption{Values of the real and imaginary parts of the amplitudes for five
lowest partial waves deduced from fits to the ANKE \pppi\ and \nppppi\
measurements at 353~MeV. Also shown are the ratios of the imaginary to real
parts of the amplitudes that have been freely fitted. The other three ratios
are fixed by the Watson theorem.} \label{tab:fit}
\end{table}

%
%

\section{Conclusions}
\label{sec7}

We have studied the $np\to \{pp\}_{\!s}\pi^-$ reaction in the vicinity of
353~MeV per nucleon using a polarized deuteron beam incident on a polarized
hydrogen target cell. Although the primary aim was the measurement of the
transverse spin-correlation coefficient $A_{x,x}$, the proton analyzing power
$A_y^p$ was also determined by averaging over the deuteron beam
polarizations. The consistency of these latter results with those obtained
earlier with a proton beam~\cite{DYM2012} suggests that many systematic
effects are indeed under control. This belief is reinforced by the values of
the differential cross section that were extracted from the new
double-polarized data. Although the luminosity could not be reliably
evaluated for these data, the shape of the cross section agreed well with our
published data~\cite{DYM2012} and, in particular, showed the small
forward-angle peak that was absent from the TRIUMF pion-production
results~\cite{DUN1998}, though there are indications of it in their pion
absorption data~\cite{HAH1996}.

In the earlier experiment~\cite{DYM2012} the spectator protons were detected
in the silicon tracking telescopes and, in order to control the background,
only spectators that had sufficient energy to pass completely through the
first silicon wafer were accepted. However, with increased understanding of
the detector timing information, in a reanalysis of the old data, events have
been retained where the proton stopped in the first STT layer. These stopping
events give results that are consistent with those of the previous set and
the reanalysis has doubled the $A_y^p$ statistics. These events have also
allowed us to investigate the effect of making a tighter selection on the
diproton excitation energy. Taking a cut at 1.5~MeV, as used at
TRIUMF~\cite{HAH1999}, gives very similar $A_y^p$ results to our original
3~MeV cut.

The combined \pppi\ and \nppppi\ data sets have been used in the partial wave
analysis presented here and no attempt has been made to include any effects
associated with the breaking of isospin invariance. For example, although
$\delta_{^{3\!}P_0}$ is suggested to be $-14.8^{\circ}$ in $pp$ elastic
scattering, the corresponding figure in the $np$ case is
$-16.7^{\circ}$~\cite{ARN2007}. There are also effects that must arise from
the pion mass differences and this is one reason for insisting on a direct
measurement of $A_{x,x}$ rather than trying to deduce its value from
independent measurements of the \pppi\ and \nppppi\ differential cross
sections and relying on Eq.~(\ref{Axx_rel}).

We can see from Fig.~\ref{fig:pwaAll4} that the three  partial wave solutions
of Table~\ref{tab:fit} can all describe reasonably well the measured values
of the differential cross section, the analyzing power, and the transverse
spin correlation in the \nppppi\ reaction. Furthermore, the \pppi\
observables~\cite{TSI2012} are also well reproduced. However, the predictions
for $A_{x,z}$ in Fig.~\ref{fig:pwaAll4}d are radically different, especially
between solution-2 and the other two. Hence even a low statistics measurement
of this parameter would be sufficient to resolve some of the residual
ambiguities. This would require the rotation of the proton polarization into
the longitudinal direction, which could be achieved for a proton beam through
the use of a Siberian snake. It is hoped that such an experiment will be
carried out at ANKE~\cite{DYM2013} using the polarized deuterium target that
was successfully commissioned in 2012. However, it should be noted that at
the Indiana cyclotron a longitudinally polarized hydrogen target was achieved
through the use of Helmholtz coils~\cite{RIN2000}. This approach would
require significant extra development work at COSY.

Are there any theoretical indications as to which of the three solutions of
Table~\ref{tab:fit} is to be preferred? Due to the strong coupling between
the $^{3\!}S_1$ and $^{3\!}D_1$ partial waves, one cannot rely on using the
Watson theorem to deduce the phases of the $M_p^S$ and $M_p^D$ amplitudes.
However, we would naively expect that the phases of the solutions that we
have found should not differ drastically from the corresponding phases of
elastic nucleon-nucleon scattering. It is interesting to note that the phases
of the $p$-wave production amplitudes evaluated within $\chi$PT stay
relatively close to the elastic phases, in spite of the coupled-channel
dynamics~\cite{BAR2009}. It is important to note here that, although the
Watson theorem suggests that the real production amplitude should acquire the
elastic phase, one does not know if the ``bare'' amplitude is positive or
negative. As a consequence there is a 180$^\circ$ ambiguity or,
alternatively, it is only the tangent of the phase that is relevant.

Turning now to the results in Table~\ref{tab:fit}, one finds that
$(\textrm{Im}(M_p^S)/\textrm{Re}(M_p^S),
\textrm{Im}(M_p^D)/\textrm{Re}(M_p^D)) = (-0.44,-1.32),\ (0.02,-0.48),\
\textrm{and}\ (0.29,-0.53)$ for the three solutions. These are to be compared
with the nucleon-nucleon phase-shift analysis values of
$(\tan\delta_{^{3\!}S_1},\tan\delta_{^{3\!}D_1})=(0.03,-0.46)$~\cite{ARN2007},
and to the values from the theoretical analysis of
$(0.04,-0.61)$~\cite{BAR2009}. Although this theoretical calculation does not
coincide exactly with the elastic phases, it is certainly much closer to
solution-2 than solution-3. Specifically, the difference of 0.13 in
$\tan\delta_{^{3\!}D_1}$ between theory and solution-2 corresponds to a phase
difference of only $5^{\circ}$ whereas the difference for solution-3 is
already $14^{\circ}$. There is therefore a distinct preference against
solution-1 and possibly in favor of solution-2. However it is difficult to
quantify what emphasis should be placed on these theoretical arguments and a
direct measurement of $A_{x,z}$ is required to clarify the situation.

We have made several assumptions in the partial wave fittings, especially by
neglecting the interferences between the $s$- and $g$-waves and, more
contentiously, between the $p$- and $f$-waves. Their inclusion within some
model might change slightly the fit parameters in Table~\ref{tab:fit}.
Nevertheless we believe that the solutions achieved are now sufficiently
robust that they can be used in the framework of a modified Chiral Perturbation Theory
to achieve some of the goals outlined in the introduction.

%
%
\begin{acknowledgments}
We are grateful to other members of the ANKE Collaboration for
their help with this experiment and to the COSY crew for
providing such good working conditions, especially of the
polarized beam. The discussions with C.~Hanhart on this problem
over several years have been very illuminating. The work was
supported by the COSY FFE program, STFC ST/J000159/1 grant,
 and the Shota Rustaveli National Science Foundation Grant 09-1024-4-200.
\end{acknowledgments}
%
%

\end{document}